\documentclass[9pt,twocolumn,twoside,lineno,pdftex]{pnas-new}

\usepackage[normalem]{ulem}
\usepackage{microtype}
\usepackage[charter]{mathdesign}
\usepackage{color}         

\usepackage{bm}
\usepackage{natbib}
\usepackage{placeins}
\usepackage{tikz}

\usetikzlibrary{shapes,shadows,arrows}
\usepackage[americanvoltages, europeancurrents, americanresistors, americaninductors,oldvoltagedirection]{circuitikz}
\definecolor{mygreen}{rgb}{0,0.5,0}
\definecolor{mybrown}{RGB}{165,42,42}
\definecolor{mymagenta}{RGB}{255,0,255}
\definecolor{myyellow}{RGB}{255,255,0}
\definecolor{mycyan}{RGB}{0,255,255}
\usepackage[version=4]{mhchem}

\templatetype{pnasresearcharticle} 

\title{Low-temperature emergent neuromorphic networks with correlated oxide devices}

\author[a,1]{Uday S. Goteti}
\author[a,1]{Ivan A. Zaluzhnyy} 
\author[b]{Shriram Ramanathan}
\author[a,2]{Robert C. Dynes}
\author[a,2]{Alex Frano}

\affil[a]{Department of Physics, University of California, San Diego, California 92093, USA}
\affil[b]{School of Materials Engineering, Purdue University, West Lafayette, IN 47907, USA}

\leadauthor{Goteti} 

\significancestatement{Designing neuromorphic hardware for cryoelectronics is an important area of research as the field of computing paradigms beyond CMOS progresses. Superconductivity and metal-insulator transitions are two of the most celebrated emergent, collective properties found in quantum materials such as strongly correlated oxides. Here, we present simulations of artificial neural networks that can be designed by combining superconducting devices (e.g. Josephson junctions) with Mott metal-insulator transition-based tunable resistor devices. Our simulations show that (a) neurons and synapses can be seamlessly created, (b) their functions can be tuned via learning, and (c) controlling disorder by incorporating light ions enables exponential multiplicity of states. The results open up new directions for incorporating emergent behavior seen in condensed matter into hardware design for artificial intelligence.}

\authorcontributions{U.G. and I.Z. performed the simulations under guidance of S.R., R.C.D., and A.F. All authors wrote the paper. R.C.D. and A.F. led the project.}
\equalauthors{\textsuperscript{1}These authors contribted equally to this work.}
\correspondingauthor{\textsuperscript{2}To whom correspondence should be addressed. E-mail: rdynes@ucsd.edu, afrano@ucsd.edu}


\begin{abstract}
\nolinenumbers
Neuromorphic computing--which aims to mimic the collective and emergent behavior of the brain’s neurons, synapses, axons, dendrites--offers an intriguing, potentially disruptive solution to society's ever-growing computational needs. Although much progress has been made in designing circuit elements that mimic the behavior of neurons and synapses, challenges remain in designing networks of elements that feature a collective response behavior. We present simulations of networks of circuits and devices based on superconducting and Mott-insulating oxides that display a multiplicity of emergent states that depend on the spatial configuration of the network. Our proposed network designs are based on experimentally known ways of tuning the properties of these oxides using light ions. We show how neuronal and synaptic behavior can be achieved with arrays of superconducting Josephson junction loops, all within the same device. We also show how a multiplicity of synaptic states could be achieved by designing arrays of devices based on hydrogenated rare-earth nickelates. Together, our results demonstrate a new research platform that utilizes the collective macroscopic properties of quantum materials to mimic the emergent behavior found in biological systems.
\end{abstract}

\doi{\url{https://doi.org/10.1073/pnas.2103934118}}

\begin{document}
\nolinenumbers
\maketitle
\thispagestyle{firststyle}
\ifthenelse{\boolean{shortarticle}}{\ifthenelse{\boolean{singlecolumn}}{\abscontentformatted}{\abscontent}}{}

\begin{figure*}[!htb]
    \centering
    \includegraphics[width=0.75\linewidth]{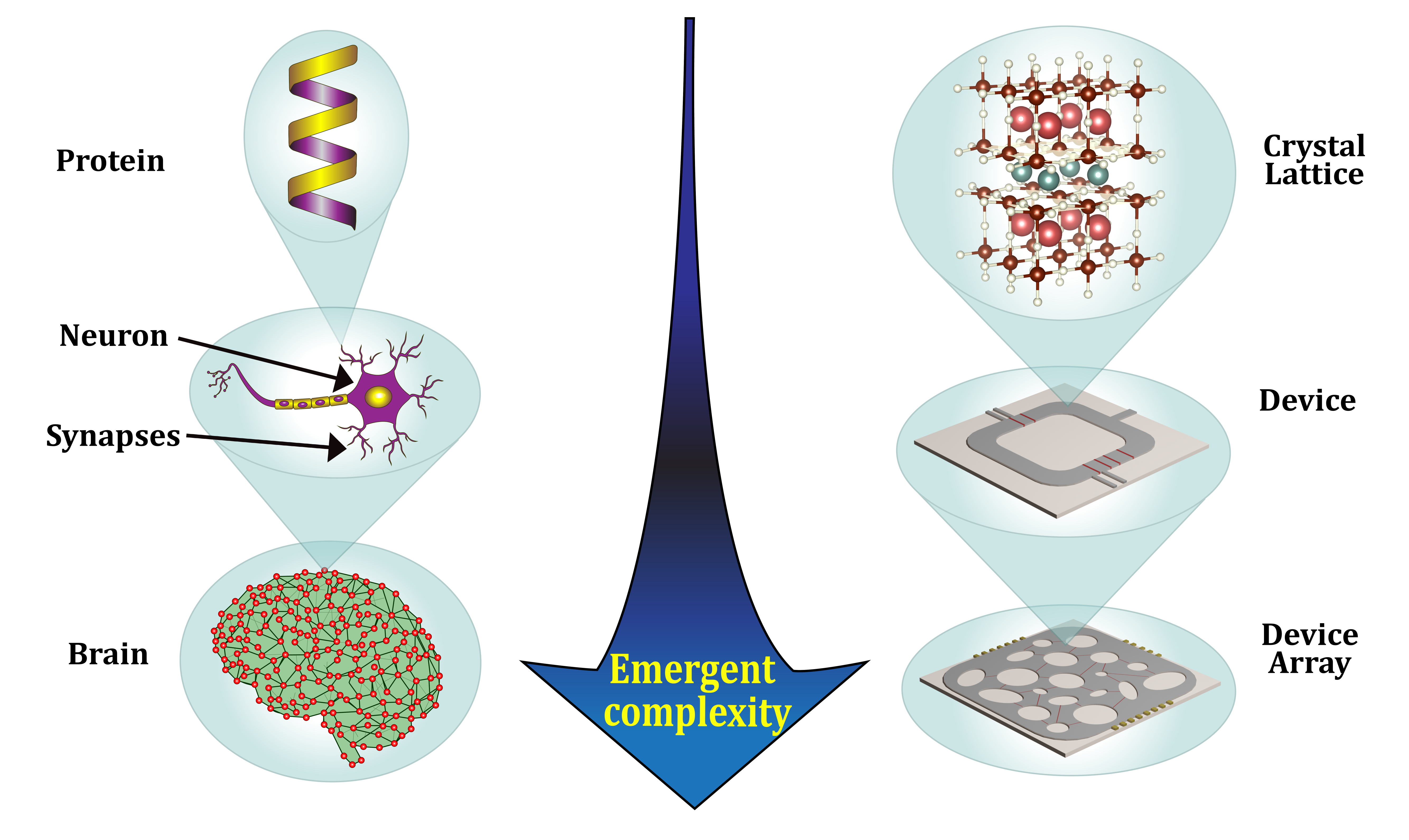}
    \caption{A comparison of the emergent behavior that arises in biological systems, including simple living organisms (left), and artificial (right) devices. Disorder and randomness play roles at all length scales. In the case of correlated oxides (right), disorder in the lattice can yield different macroscopic properties that can be used to make devices by controlled light-ion modifications. Moreover, a randomly designed network of said devices can yield exponentially more complex, emergent responses. }
    \label{Hierarchy}
\end{figure*}

\nolinenumbers
\dropcap{E}mergent behavior, defined as when a system's collective behavior is different than that of its individual constituents, is widespread and consequential in nature. The brain, for example, is made up of proteins, tissue and flowing chemicals and charges but functions collectively over a long spatial range as a uniquely powerful and energy-efficient machine. Remarkably, its constituent elements and global architecture feature staggering amounts of seeming randomness and disorder (Fig. \ref{Hierarchy}), yet has evolved to be capable of astounding computational functionalities: in many ways still more impressive and energy-efficient than semiconductor-based computers. Analogously, quantum materials, such as strongly correlated systems~\cite{Keimer2017}, display collective macroscopic behavior such as superconductivity~\cite{Keimer2015} and metal-insulator transitions~\cite{Imada1998}. These macroscopic collective responses emerge from microscopic quantum mechanical interactions. As a result, brain-inspired computing paradigms--known  broadly as neuromorphic~\cite{Mead1990, Li_2018}--based on these quantum materials are prominent in the goals of various research efforts to explore and hopefully spawn the next technological revolution~\cite{ramanathan_2018,delValle2018_JAP,Shi2013, Shi2014,Rosillo2020}.

Natural neural networks in animal brains are comprised of neurons that are interconnected by synapses. Neurons are capable of integrating charge and releasing them at critical thresholds referred to as action potentials. Synapses can amplify or decrease the signal strength by chemical or electrical pathways~\cite{tibbetts_principles_2013}. Information is encoded temporally in such networks and represents a distinct paradigm from traditional digital electronics. Synapses store memory and can dynamically adjust their weight in response to the time intervals between neuronal stimuli (known as time-dependent plasticity). Neuromorphic hardware networks therefore aspire to capture the key features found in the nervous
system such as periodic trains of spiking signals; multi-state memory that can be programmed incrementally as well as in a time-dependent manner. Materials that can host diverse electronic structures and / or present non-linear electrical characteristics often are promising candidate systems to explore as building blocks for neuromorphic networks. Further, to emulate the complexity of natural networks, having multiple control knobs via ionic or electronic inputs to tune the order parameter in neuromorphic devices is desirable.

Neuromorphic computing architectures based on correlated transition metal oxides could offer a flexible, low-consumption alternative to von Neumann architectures. The key challenge is to generate flexible material response properties that can harbor multiple states to allow flexibility and an architecture for computation and memory to operate in parallel. Correlated oxides offer a platform to explore high-density and multi-state memory because of the opportunity to apply their multiple phases including superconductivity, magnetism, and metal-insulator transitions. While low-temperature CMOS and cryo-cooling have long been an active area of research~\cite{Gubser1997, Kirschman1990, vanDijk2020}, cryoelectronic neuromorphic device technologies based on superconducting materials are gaining interest rapidly due to their unique advantages in power efficiency and in flexibility to generate spiking neuron-like behavior~\cite{Schneider2018, Shainline2017,shainline2019superconducting,toomey2019design,segall2017synchronization,li2012simulating,schneider2018ultralow,harada1991artificial,bozbey2018single,cheng2018spiking,schneider2020fan,zhang2020phase}.

In this paper, we report results of simulations that demonstrate new classes of low temperature artificial neural networks that arise from designing controlled-disorder in devices based on correlated oxides. We utilize two archetypal properties for this purpose, namely superconductivity and metal–insulator transitions, with the common theme of creating controlled disorder with light-ions incorporation.  

First, we will discuss how lattice disorder in superconducting YBa$_2$Cu$_3$O$_7$ (YBCO) induced by helium ion implantation can be used to fabricate arrays of superconducting Josephson junction loops that yield an exponential multiplicity of coherent states. This emergent multiplicity can further evolve by randomizing the array’s spatial geometry. The disordered superconducting loops are fast and allow multiple states that have a transient nature with low energy consumption. Second, we discuss how arrays of hydrogen-ion incorporated rare-earth nickelate devices, each of which are known to individually render synaptic behavior, can also yield a multiplicity of states that rely on the spatial configuration and design of the array. These responses are slower than the superconducting loops, but more stable in the various states and are more memory centric. While the different oxide material systems possess different microscopic behaviors, we illustrate that the flexibility of these two case-study systems allow high-density architectures with configurational randomness that yield many possible states that could mimic the animal brain’s notable random-to-collective behavior.

Moreover, our simulations outline a broad research effort to harness the properties of strongly correlated electron systems to produce arrays of disordered individual brain-inspired elements that collectively evolve to render emergent functionalities. These quantum materials feature a large range of systems with ``binary'' states that can be employed, and we show how these can be configured to construct neurons and synapses on the same device. While the two examples we present are near the ends of the spectra of speed, energy, and stability, the range of properties in oxides allow a wide range of speed, power consumption and volatility, i.e., long-term stability or transient volatility. With the wealth of oxides currently being studied throughout a large community, we expect many more systems following the architectural framework that this paper discusses.

This paper is organized as follows. First, we summarize the experimentally known properties that enable our neuromorphic simulations--superconductivity and metal-insulator transitions--particularly focusing on the effect that light ions have on them. Then, we discuss how superconducting Josephson junction loops could render neuronal behavior in copper-oxide devices. Next, we discuss how synaptic behavior can be achieved in two platforms. Finally, we look outwards by presenting examples of connectivity between the devices we propose and other material platforms.

\section*{Quantum Material Platforms}

The emergent neuromorphic models we outline below are founded on two archetypal collective electronic phases that occur in transition metal oxides, namely high-temperature superconductivity and Mott metal-insulator transitions. Specifically, we focus on copper oxides and rare-earth nickelates, respectively, although in the case of metal-insulator materials, there are several other oxides that can be used. Importantly, there is ample experimental evidence showing that both of these phases can be tuned by using light ions such as H, Li, and/or He. Thus, before describing our neuromorphic models, we first summarize the effect of light ions.

The superconductivity and normal-state transport properties of high-Tc superconductor YBa$_2$Cu$_3$O$_7$ (YBCO) have been shown to be highly sensitive to ion irradiation in various studies \cite{clark1987effects,valles1989ion,basov1994disorder,lesueur1993ion}. Particularly, the material undergoes a continuous transition from metallic behavior in the normal state to insulating behavior as the irradiation dose of ions is increased \cite{valles1989ion}. These effects are explained as due to induced disorder by the ion bombardment rather than doping as evident in previous reports \cite{clark1987effects,valles1989ion,cybart2014comparison}. 

Therefore, planar superconducting tunnel junctions have been constructed by using modern focused He-ion beams with a beam diameter, and therefore the tunnel barrier width, of $\sim$500\,pm \cite{cybart2015nano}. With the appropriate choice of ion dosages, Superconductor-Insulator-Superconductor and Superconductor-Normal-Superconductor junctions can be constructed. The ability to produce Josephson tunnel junctions of arbitrary shape and size with tunable junction properties such as critical current density opens a possibility to explore novel neuromorphic devices.

Superconducting Josephson tunnel junctions, formed using damage induced by a focused ion beam on YBCO thin films \cite{cybart2015nano}, can be viewed as relaxation oscillators. When a current through the tunnel barrier exceeds the critical current $I_C$ of the junction, the phase difference of the superconducting order parameter oscillates with a frequency proportional to the voltage developed across the junction. These oscillations in YBCO tunnel junctions using focused ion beams \cite{cybart2015nano} can be sufficiently damped to produce non-hysteretic current-voltage characteristics, and therefore can produce spiking voltage, each corresponding to single-flux quantum vortex with a total change in the phase difference of $2\pi$ \cite{likharev1991rsfq}.

The relaxation oscillations and spiking characteristics of Josephson junctions have been of interest in different artificial neural networks \cite{schneider2020synaptic,shainline2017superconducting,crotty2010josephson,schneider2018ultralow}. Furthermore, alternative devices to generate spiking neuron functionality has also been proposed using superconducting nanowire-based devices \cite{cheng2018spiking,toomey2019design}. However, the disordered YBCO junctions can be particularly suitable for superconducting neurons and disordered array synaptic network elements as described below. Specifically, a single array can be configured to achieve both neuron and synapse operation. Additionally, several such neurons can be connected through the proposed disordered array synaptic networks to form recurrent neural networks with all the comprising elements evolving together. Furthermore, such networks can be interconnected through similar disordered arrays to form a hierarchical recurrent network similar to a biological brain \cite{uday2020disordered}.

In the case of the rare earth nickelates, \ce{RNiO3}, one can influence the metal-insulator transition (MIT) by electron doping of the material independent of temperature~\cite{Zuo2017,Yoo2018,Kotiuga2019a}. The addition of one extra electron to the nickel ion induces a large Mott–Hubbard splitting due to strong Coulomb interaction between the electrons on the same $e_g$ orbital of Ni.
As a result, a large $\sim$3\,eV electronic band gap opens between the highest occupied orbitals of O and lowest unoccupied orbitals of Ni, which leads to a dramatic increase of resistance \cite{Shi2014}.

Experimentally this effect was demonstrated by doping \ce{SmNiO3} and \ce{NdNiO3} with hydrogen \cite{Shi2013,Shi2014,Zhou2016,Zuo2017,Zhang2018} and lithium ions \cite{Shi2014,Sun2018}.
In these works, doping with hydrogen was performed by depositing Pd or Pt on top of the nickelate film and subsequent one-time annealing in hydrogen atmosphere.
During this process, the hydrogen molecules \ce{H2} catalytically dissociate into individual atoms which diffuse into the nickelate film.
Homogeneously doping the material with hydrogen can yield a $>10^8$ non-volatile increase in electrical resistance \cite{Shi2014}.
Recently, x-ray absorption and diffraction experiments have shown that the main mechanism of such colossal resistance change during hydrogenation lies in bringing in an extra electron to the nickel ions, while the changes in the \ce{SmNiO3} crystal structure due to doping are very small \cite{Zhang2020a}. Since the electronic phase transition and concurrent resistance tuning is independent of temperature, it is possible to couple nickelate synapses with superconducting junctions to create neural networks.

\section*{Simulations of Loop Neuron Structures with Ion-Damaged Tunnel Junctions}

\begin{figure*}[!htb]
    \centering
    \begin{tikzpicture}
        \fill[red!50!blue] (0,0) -- (0,5) -- (4,5) -- (4,0) -- (0,0);
        \fill[white] (0.5,0.5) -- (0.5,4.5) -- (3.5,4.5) -- (3.5,0.5) -- (0.5,0.5);
        \fill[red!50!blue] (1.75,5) -- (1.75,6) -- (2.25,6) -- (2.25,5) -- (1.75,5);
        \fill[red!50!blue] (-0.5,2.2) -- (-0.5,2.3) -- (0,2.3) -- (0,2.2) -- (-0.5,2.2);
        \fill[red!50!blue] (-0.5,2.7) -- (-0.5,2.8) -- (0,2.8) -- (0,2.7) -- (-0.5,2.7);
        \fill[red!50!blue] (4,0.7) -- (4,0.8) -- (4.5,0.8) -- (4.5,0.7) -- (3,0.7);
        \fill[red!50!blue] (4,1.2) -- (4,1.3) -- (4.5,1.3) -- (4.5,1.2) -- (3,1.2);
        
        \draw[black] (-0.5,2.75) node[anchor=north east]{Spiking input};
        \draw[black] (4.5,1.25) node[anchor=north west]{Spiking output};
        \draw[black] (2,6.5) node[anchor=north]{DC current input};
        
        \draw[yellow,very thick] (0,2.5) -- (0.5,2.5);
        
        \draw[yellow,very thick] (3.5,1) -- (4,1);
        \draw[yellow,very thick] (3.5,1.5) -- (4,1.5);
        \draw[yellow,very thick] (3.5,2) -- (4,2);
        \draw[yellow,very thick] (3.5,2.5) -- (4,2.5);
        \draw[yellow,very thick] (3.5,3) -- (4,3);
        \draw[yellow,very thick] (3.5,3.5) -- (4,3.5);
        \draw[yellow,very thick] (3.5,4) -- (4,4);
    \end{tikzpicture}
    \caption{\color{black}Schematic of a spiking integrate-and-fire neuron with a large superconducting integrating loop, with Josephson junctions generated using focused He-ion tunnel barriers. Ion-damaged barriers are shown in yellow and YBCO film shown in purple.}
    \label{neuron}
\end{figure*}
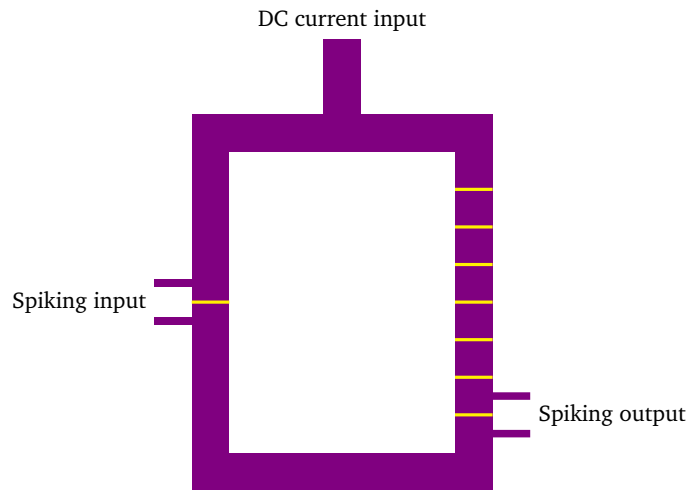

The spiking behavior in neurons and synapses are realized through generation and propagation of single-flux quanta in overdamped Josephson junctions and superconducting loops incorporating these junctions respectively. The flux shuttle \cite{fulton1973flux} structure describes the vortex dynamics and the corresponding Josephson phase oscillations in junctions and superconducting loops. Such vortex motion in superconducting loops is being employed in digital computing systems such as single-flux quantum logic circuits \cite{likharev1991rsfq}. Therefore, several of the well-established design and hardware ideas used in those systems may also be useful in developing neurons and synapses presented here. An integrate-and-fire neuron behavior can be realized by appropriately designing these loops and junctions. An example is shown in Fig. \ref{neuron} that produces a spiking voltage output when the current in the integrating superconducting loop exceeds the critical current of the output tunnel junction. The operation of the structure in Fig. \ref{neuron} is discussed in detail in \cite{uday2020disordered}, where the neuron is shown to exhibit firing dynamics of a leaky integrate-and-fire neuron.

\begin{figure*}[!htb]
    \centering
    \includegraphics[width=0.99\linewidth]{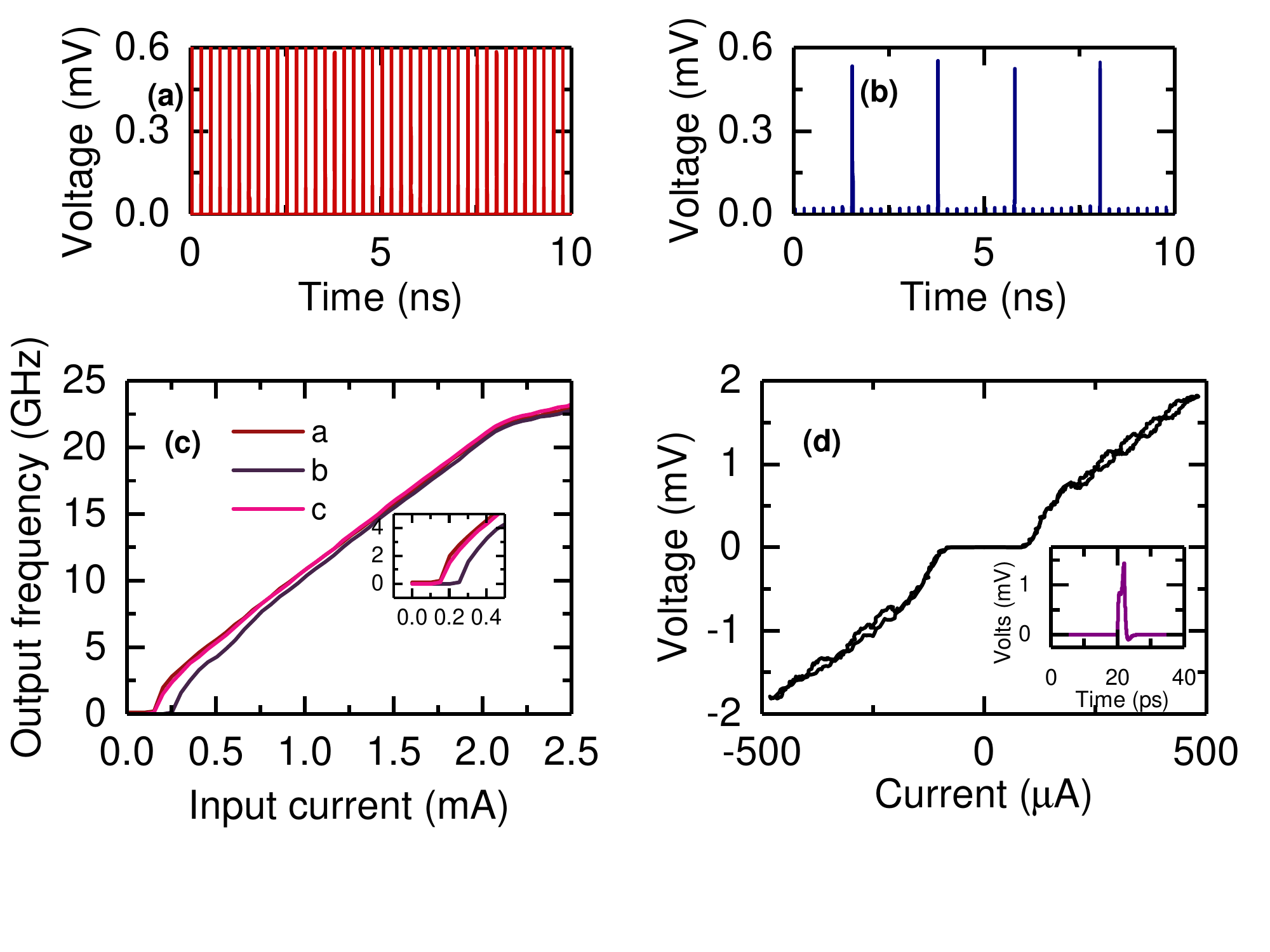}
    \caption{Simulation results of {\color{black} the spiking integrate-and-fire} neuron shown in Fig. \ref{neuron}. Critical currents of all the junctions in the loop are chosen to be 100 $\mu$A, and the total inductance of the loop is 100 pH. (a) Spike-train of constant frequency 
    {\color{black}of 250 MHz applied} to the $Spiking$ $input$ terminal of Fig. \ref{neuron}. (b) Output spikes across the $Spiking$ $output$ terminal of Fig. \ref{neuron} fired by the neuron as the current in the integrating loop reached a threshold. {\color{black}A spiking frequency of the output is 25 MHz yielding a threshold of 10.} (c) Input current versus output frequency of the superconducting neurons of different sizes and number of junctions in the stack. The behavior represents that of a leaky integrate-and-fire neuron. Inset with zoomed-in view of input current versus output frequency to show that different threshold behaviors can be achieved by appropriately designing the size of the loop and the junction stack. a - 10 junctions in the loop with a loop inductance of 100 pH. b - 10 junctions in the loop with a loop inductance of 200 pH. c - 8 junctions with a loop inductance of 100 pH. {\color{black}(d) Typical I-V characteristics of the Josephson junction used in the integrate-and-fire neuron. The junction is excited with a current pulse above its critical current to generate a single-flux quantum spike. Inset shows example of a single spike generated across Josephson junction by exciting the junction with a short current pulse of 300 $\mu$A.}}
    \label{neuron-sim}
\end{figure*}

The parameters of the neuron, such as the maximum current in the integration loop before the neuron fires, the time constants of the decay of the loop current, and the resting potential can be controlled using the size of the loop and the parameters of junctions in it as demonstrated by the simulation results in Fig. \ref{neuron-sim}. An example of the current-voltage characteristics of one of the Josephson junctions used in the neuron is shown in Fig. \ref{neuron-sim}e. The junctions are overdamped, thereby producing single-flux quantum voltage spikes when excited with a current pulse with magnitude above their critical currents as shown in Fig. \ref{neuron-sim}f. Therefore, as the current in the integration loop increases beyond the junction critical current, a spiking output is produced. Different neurons in a neural network can be designed accordingly by simply adjusting the loop size and the ion damage of the tunnel barrier that defines the junction critical current. A spike-train of constant frequency is applied at the input of the neuron as shown in Fig. \ref{neuron-sim}a, which switch the input junction and enter the integration loop. The current in the loop reaches a threshold defined by the critical currents of the output junction stack as the input spikes enter the loop. The threshold can be dynamically varied within a range defined by the $DC$ $current$ $input$ shown in Fig. \ref{neuron}. The spiking output of the neuron is shown in Fig. \ref{neuron-sim}b, as the threshold is gradually varied. As the $DC$ $current$ $input$ is increased, the output spike frequency increases exhibiting a current versus frequency behavior of that of an ideal integrate-and-fire neuron. Results of three different geometries and junction parameters are shown in Fig. \ref{neuron-sim}c. Additionally, the maximum threshold of the neuron is also defined by the physical design of the structure as shown in the inset Fig. \ref{neuron-sim}d.

\section*{Simulations of Disordered Array Synaptic Networks}

The relaxation oscillator structures that define the neuron behavior of the previous section can be connected to other neurons via the following synaptic systems.

\subsection*{Cuprate-Based Synapse Arrays}

The cuprate-based neurons of the previous section can also be useful for producing synaptic behavior in disordered arrays as shown in Fig.~\ref{dis_array}, highlighting the singular scenario where one device can act as a neuron or synapse. 
While the neuron structure has a varying threshold and firing behavior that depends on the input magnitude (i.e., the number of input spikes or the total DC current input), a disordered array structure that is decoupled from the input spike timing or rate can exhibit output behavior dependent on the input or feedback spike timing (spike timing dependent plasticity), spiking rate (rate dependent plasticity) and the total magnitude (i.e., the number of vortices or spikes).
Furthermore, the large non-volatile memory of the array comes from the various configurations of supercurrents within different loops. This memory state updates when vortices enter and leave the array with every change in input or output thereby allowing learning behavior. 
If each of the loops in the array is restricted to allow at least one vortex in it, then the total number of memory configurations possible can be a maximum of $3^n$, where $n$ is the number of loops. Note that this can be a significantly higher number if the loops can accommodate more than one vortex. However, any spatial symmetry in the pattern of arrays reduces this number. So a disordered array has the most possible number of configurations. The array has input and output terminals (shown in blue in Fig.~\ref{dis_array}) for spiking signals and feedback terminals (shown in black in Fig. \ref{dis_array}). Therefore, it behaves as a collective synaptic network between all the input and output neurons connected to the array. This approach presents an alternative architecture to artificial neural networks that is described in detail in a previous paper \cite{uday2020disordered}. In summary, this disordered array approach is defined by the small-world or random fully recurrent neural network as opposed to largely feed-forward and regular neural network architectural approach popular in artificial neural networks. Consequently, the disordered array approach can be a highly scalable and power efficient approach \cite{watts1998collective}.

\begin{figure*}[!thb]
    \centering
    \begin{tikzpicture}
        \draw[blue] (0,1) -- (-1,1) node[anchor=north west]{i1};
        \draw[blue] (0,2) -- (-1,2) node[anchor=north west]{i2};
        \draw[blue] (0,3) -- (-1,3) node[anchor=north west]{i3};
        \draw[blue] (0,4) -- (-1,4) node[anchor=north west]{i4};
        \draw[blue] (5,1) -- (6,1) node[anchor=north east]{o4};
        \draw[blue] (5,2) -- (6,2) node[anchor=north east]{o3};
        \draw[blue] (5,3) -- (6,3) node[anchor=north east]{o2};
        \draw[blue] (5,4) -- (6,4) node[anchor=north east]{o1};
        \draw[black] (1,5) -- (1,6) node[anchor=north west]{b1};
        \draw[black] (2,5) -- (2,6) node[anchor=north west]{b2};
        \draw[black] (3,5) -- (3,6) node[anchor=north west]{b3};
        \draw[black] (4,5) -- (4,6) node[anchor=north west]{b4};
        
         \fill[red!50!blue] (0,0) -- (0,5) -- (5,5) -- (5,0) -- (0,0);
         \draw[yellow,very thick] (0.75,1) -- (1.65,0.8);
         \draw[yellow,very thick] (0,1) -- (0.5,1);
         \draw[yellow,very thick] (0.65,1.1) -- (0.9,2);
         \draw[yellow,very thick] (1.3,2) -- (3,3);
         \draw[yellow,very thick] (1.1,2.8) -- (3,2.7);
         \draw[yellow,very thick] (1.4,4) -- (2.2,3.1);
         \draw[yellow,very thick] (1,3.9) -- (0,3.9);
         \draw[yellow,very thick] (0,3) -- (0.85,2.9);
         \draw[yellow,very thick] (2,0.5) -- (3.6,0.7);
         \draw[yellow,very thick] (4.2,2.9) -- (5,2.4);
         \draw[yellow,very thick] (3.8,0.8) -- (5,1);
         \draw[yellow,very thick] (2.5,3.7) -- (3.2,4.1);
         \draw[yellow,very thick] (4,4) -- (5,4);
         \draw[yellow,very thick] (2.7,2.3) -- (3.8,1.9);
         \draw[yellow,very thick] (3.8,2) -- (4,2.8);
         \draw[yellow,very thick] (4,1.7) -- (5,1.5);
        
        \fill[white] (2.5,3) ellipse (0.5cm and 1.2cm);
        \fill[white] (0.5,1) ellipse (0.3cm and 0.2cm);
        \fill[white] (1,2) ellipse (0.4cm and 0.4cm);
        \fill[white] (3.5,0.9) ellipse (0.6cm and 0.4cm);
        \fill[white] (1.8,0.6) ellipse (0.5cm and 0.5cm);
        \fill[white] (0.8,4) ellipse (0.7cm and 0.6cm);
        \fill[white] (1,2.8) ellipse (0.25cm and 0.25cm);
        \fill[white] (3.6,4) ellipse (0.4cm and 0.6cm);
        \fill[white] (4.2,2.9) ellipse (0.35cm and 0.3cm);
        \fill[white] (4,2) ellipse (0.4cm and 0.45cm);
    \end{tikzpicture}
    \caption{\color{black}Collective synapse with 4 inputs and 4 outputs employing a disordered array of loops of different sizes connected through Josephson junctions. Flux quanta can be stored in the loops in the form of circulating supercurrents, with junctions allowing transport of flux between loops. $i1,i2,...$ are the input terminals, $o1,o2,...$ are the output terminals and $b1,b2,...$ are the feedback terminals coupled to the outputs.}
    \label{dis_array}
\end{figure*}
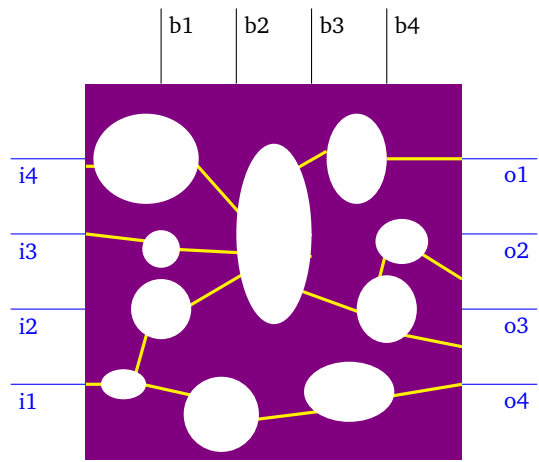

\begin{figure*}
    \centering
    \begin{tikzpicture}
        \fill[red!50!blue] (0,0) -- (0,3) -- (3,3) -- (3,0) -- (0,0);
        \fill[red!50!blue] (-1.5,-3) -- (-1.5,0) -- (4.5,0) -- (4.5,-3) -- (-1.5,-3);
        \fill[red!50!blue] (1.4,3) -- (1.6,3) -- (1.6,3.5) -- (1.4,3.5) -- (1.4,3);
        \fill[red!50!blue] (-1.5,0) -- (-2,0) -- (-2,-0.2) -- (-1.5,-0.2) -- (-1.5,0);
        \fill[red!50!blue] (4.5,0) -- (5,0) -- (5,-0.2) -- (4.5,-0.2) -- (4.5,0);
        
        \draw[black] (1.5,4) node[anchor=north]{Spiking input};
        \draw[black] (-2,0) node[anchor=north east]{Current feedback 1};
        \draw[black] (5,0) node[anchor=north west]{Current feedback 2};
        
        \draw[yellow,very thick] (0,1.5) -- (0.6,1.5);
        \draw[yellow,very thick] (2,1.5) -- (3,1.5);
        \draw[yellow,very thick] (-1.5,-1.5) -- (-1.2,-1.5);
        \draw[yellow,very thick] (1.2,-1.5) -- (2,-1.5);
        \draw[yellow,very thick] (3.5,-1.5) -- (4.5,-1.5);
        
        \fill[white] (1.3,1.5) ellipse (1.1cm and 0.9cm);
        \fill[white] (0,-1.5) ellipse (1.2cm and 1.2cm);
        \fill[white] (2.8,-1.5) ellipse (0.85cm and 0.95cm);
        
        \draw[black] (1.3,1.75) node[anchor=north]{Loop 1};
        \draw[black] (-0.25,1.75) node[anchor=north]{J1};
        \draw[black] (3.25,1.75) node[anchor=north]{J2};
        \draw[black] (-0.4,-1.25) node[anchor=north]{Loop 2};
        \draw[black] (-1.75,-1.25) node[anchor=north]{J3};
        \draw[black] (0.8,-1.25) node[anchor=north]{J4};
        \draw[black] (2.75,-1.25) node[anchor=north]{Loop 3};
        \draw[black] (5,-1.25) node[anchor=north]{J5};
        
        \draw[densely dashed] (0.25,2.75) -- (2.75,2.75);
        \draw[densely dashed] (0.2,2.75) -- (0.2,0.25) node[vee]{};
        \draw[densely dashed] (2.75,2.75) -- (2.75,0.25) node[vee]{};
        \draw[densely dashed] (-1.25,-0.25) -- (4.5,-0.25);
        \draw[densely dashed] (-1.25,-0.25) -- (-1.25,-2.5) node[vee]{};
        \draw[densely dashed] (4.25,-0.25) -- (4.25,-2.5) node[vee]{};
        \draw[densely dashed] (1.5,-0.25) -- (1.5,-2.5) node[vee]{};
    \end{tikzpicture}
    \caption{\color{black}Disordered array with 3-loops representing a simplified synaptic network used in simulations and analysis. The junctions are all chosen to have different parameters and the geometry is chosen to distribute inductance asymmetrically in the network. Asymmetry ensures that the number of available memory states are maximized for a given number of loops. }
    \label{synapse}
\end{figure*}
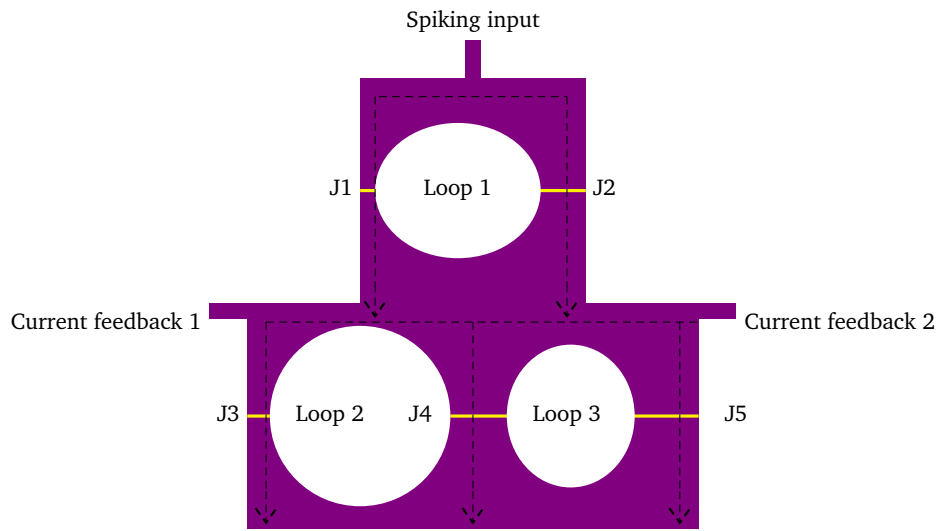

\begin{figure*}[!htb]
    \centering
    \includegraphics[width=0.99\linewidth]{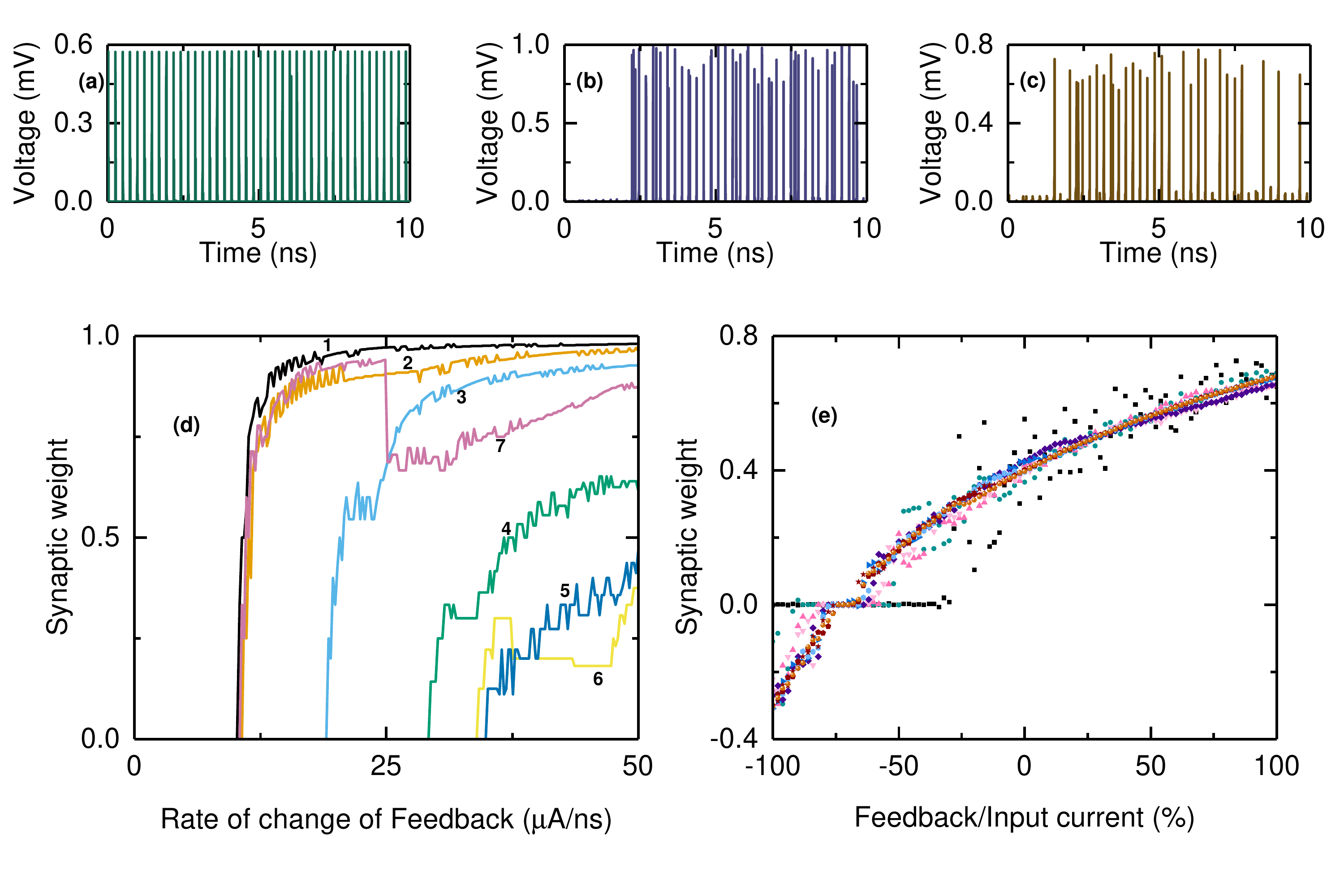}
    \caption{\color{black}Simulation results of the 3-loop synapse with 1 input and 2 outputs as shown in Fig. \ref{synapse}. {\color{black} Critical currents of the junctions are as follows: J1 (140 $\mu$A), J2 (110 $\mu$A), J3 (120 $\mu$A), J4 (100 $\mu$A), J5 (160 $\mu$A). Loop inductances are as follows: Loop 1 (15 pH), Loop 2 (45 pH) and Loop 3 (37 pH).}(a) An input spike-train of constant frequency with each spike representing single-flux-quantum applied at the input. A linearly increasing bias current is applied at $Current$ $feedback$ $1$ while the current at the $Current$ $feedback$ $2$ is zero. (b) Resulting output spike train at output 1. The output spike frequency changes non-linearly as the feedback current changes. (c) Resulting output spike train at output 2. The output spike train is different than that of output 1. (d) Synaptic weight calculated using equation (1) versus the rate of change of feedback current at $Current$ $feedback$ $1$. 1 - Only the rate of $Current$ $feedback$ $1$ is varied for a period of 10ns. 2 - Only the rate of $Current$ $feedback$ $1$ is varied with a different input frequency for a period of 10ns. 3 - Both the current feedback currents are varied linearly for a period of 10ns. 4 - The input frequency is varied along with both the current feedback rates for a period of 10ns. 5 - The input and feedback signals are linearly varied for a period of 10ns. 6 - All the signals are varied non-linearly. 7 - All the signals are varied non-linearly, but with different input signals from that of 6. (e) Synaptic weight as a function of percentage of feedback current to input current for 10 different feedback and input current values. The input current is varied within the range -10 mA and 10 mA with 10 different curves representing different current steps of 2 mA. }
    \label{synapse-sim}
\end{figure*}

The large disordered array shown in Fig.~\ref{dis_array} comprises several relaxation oscillators defined by superconducting loops which are fabricated with focused-ion beam tunnel junctions with different natural frequencies. Therefore, such large arrays of coupled oscillators describe a complex system that may be more effectively studied by examining its emergent behavior at a larger scale. To describe this in a fairly simple structure, we consider a three-loop configuration and demonstrate some of its properties that support synaptic behavior. The three-loop structure shown in Fig. \ref{synapse} is disordered with dissimilar tunnel junctions and an asymmetric geometry. Therefore, the vortex dynamics for vortices entering the array through the terminal labeled $Spiking$ $input$ and leaving through the other two loops is non-linear and highly dependent on input spiking and current biasing (feedback current) conditions. However, the mechanisms defining such dynamics can be understood by considering the current paths between any two terminals. 

The current entering the first loop divides between the two paths until it exceeds the critical current of the smaller junction. After the junction switches, the current diverts to a different path. This process is dynamic, with current paths changing with every switching event in the array. The synaptic weight corresponds to the total current between any two terminals that can be defined as input or output. Analogous to the neuron behavior, the total current and rate of change of current between any two nodes in the neural network can produce corresponding voltage spikes across the junctions at those nodes. Therefore, the synaptic weight is defined as shown in equation below:

\begin{equation}
    Synaptic \; weight = \frac{\# \; of \; spikes \; at \; the \; output \; neuron}{\# \; of \; spikes \; at \; the \; input \; neuron}
\end{equation}

The synaptic network of Fig. \ref{synapse} was resolved into lumped circuit elements and the resulting simulations of the electrical circuit are shown in Fig. \ref{synapse-sim}. An input spike train of constant frequency is sent to one of the loops and the output {\color{black}spike trains} across the outer junctions in the other two loops are shown in Fig. \ref{synapse-sim}a-c. The output spiking behavior changes with changing feedback currents and their rates of changes. \color{black} The synaptic weight can be calculated by measuring the total number of output spikes produced for a given number of input spikes as described by equation (1). The synaptic weight, and therefore the plasticity of a disordered array memory depends on several parameters such as the input spike frequency and timing, along with output spike frequency and timing through the feedback loop. This behavior is described by simulation results shown in Fig. \ref{synapse-sim}d, where the synaptic weight is plotted as a function of one of the control parameters, i.e. rate of change of $Current$ $feedback$ $1$. Different curves, labeled 1 to 7, show resulting weights as the other parameters are varied. Curve 1 shows the case where only $Current$ $Feedback$ $1$ is increased linearly resulting in the weight saturated at close to 1. A different input frequency yields different weight as in curve 2. When the second feedback current is also linearly varied, a different weight can be achieved as in curve 3. During the operation of an actual neural network, all the input and feedback current parameters can be expected to vary non-linearly, therefore yielding different synaptic weights as shown in curves 4 to 7 (details of the parameters are in the figure caption). Furthermore, all the curves also exhibit stable states in the form of saturated weights that are not affected by further changes in input/feedback parameters. The details of the actual circuit parameters and the currents applied are not relevant to the operation of the synaptic memory. This is because the resulting synaptic weights are similar when relative strengths and frequencies of the input and output signals are the identical irrespective of the actual magnitudes and frequencies of the individual signals. To demonstrate this behavior, the feedback current is varied as a function of input current for ten different input currents and the synaptic weights are shown in Fig. \ref{synapse-sim}e. Although the actual values of input and feedback currents in the simulations are varied significantly ranging between a few hundred micro-amperes to several milli-amperes, the resulting synaptic weights are the identical when the relative strengths of the signal are the same.

\subsection*{Nickelate-Based Synapse Arrays}

The distribution of doped hydrogen in \ce{RNiO3} is governed by diffusion, so the dopant is mostly concentrated in the vicinity ($\sim10^2$~nm) of the Pt or Pd electrode, through which it was annealed into the \ce{RNiO3} film \cite{Zhou2016}.
However, it has been shown that the distribution of hydrogen can be changed by applying a voltage bias between different contacts, which can be the spiking voltage pulses created by the YBCO neurons discussed previously.
The mechanism behind the redistribution of H dopants in \ce{RNiO3} films includes two main processes: drift of the charged \ce{H+} ions in the electric field and thermodiffusion of \ce{H+} ions in a temperature gradient created by Joule heating \cite{Zhang2020a, Kim2013}. However as soon as the currents flowing through the device are small, the latter effect can be neglected.
As a first approximation, one can assume the dominant mechanism  that \ce{H+} ions initially introduced into the nickelate film can later be moved across the film is by the electric field created when the voltage is applied to the contacts.  

The major effect of \ce{H+} ion redistribution is a change of the device resistance in a non-volatile way \cite{Shi2013,Shi2014,Ramadoss2018}. The resistance of such a device accumulates information on polarity, magnitude and duration of voltage pulses applied to the device in the past, which makes it a prospective candidate for creation of an artificial synapse \cite{Zhang2019a,Zhang2020,Zhang2020a}.
The synaptic weight of such an artificial synapse is inversely proportional its resistance. 

\begin{figure*}
    \centering
	\includegraphics[width = 0.6\linewidth]{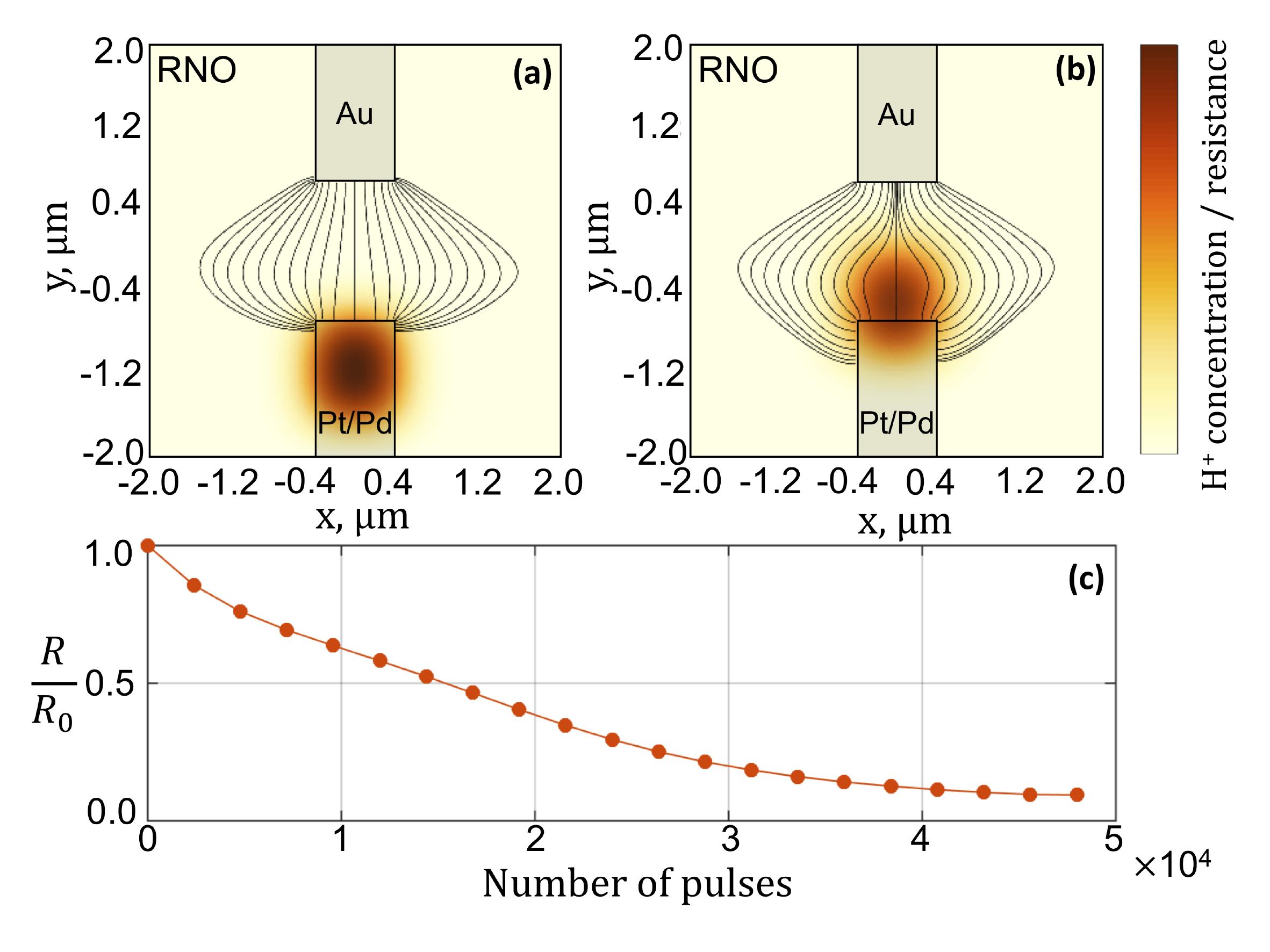}
	\caption{\label{Device1}
		(a-b) Distribution of \ce{H+} in a \ce{SmNiO3}-based memory device in low-resistance (a) and high-resistance (b) states.
		The color key corresponds to local concentration of \ce{H+} as well as electrical resistance of the media. Electric field lines between two contacts are indicated with solid black lines.
		(c) Change of the normalized resistance of the device caused by a series of electrical pulses with a magnitude 1\,mV and a duration of 1\,ns.}
\end{figure*}

A schematic of such a lateral memory device is shown in Fig.~\ref{Device1}a-b.
The voltage applied to one of the electrodes causes a drift of the implanted \ce{H+} between the Pt/Pd and Au electrodes in a two-terminal device.
Depending on \ce{H+} distribution between the electrodes, the resistance of a device can be significantly changed \cite{Zhang2020a}. 
A series of short electrical pulses applied to such a device can change its resistance, as it is shown in (Fig.~\ref{Device1}(c)). By varying the duration and magnitude of the pulses, one can change the rate at which the resistance is updated, and by that approach the learning curves typical for living neural cells \cite{Zhang2020a}. 

The resistance of such a device with any given distribution of hydrogen can be approximately estimated from a two-dimensional (2D) model we describe below.
The distribution of the current density $\bf{j}(\bf{r})$ can be found from Ohm's law $\bf{j}(\bf{r})=\sigma(\bf{r})\bf{E}(\bf{r})$, where $\bf{E}(\bf{r})=-\nabla \varphi(\bf{r})$ is an electric field, and $\sigma(\bf{r})$ is the material's conductivity.
The electrical potential $\varphi(\bf{r})$ can be found by solving by solving the current continuity equation
\begin{equation}
\label{eq3}
    \nabla (\sigma(\bf{r}) \nabla \varphi(\bf{r}))=0 \, .
\end{equation}
The conductivity $\sigma(\bf{r})$ is a function of the local concentration of hydrogen dopants $n_H(\bf{r})$, and can be approximated as \cite{Zhang2020a, Kawamoto2019}
\begin{equation}
\label{eq4}
    \sigma(\bf{r})=\sigma_0 \Big(1+\text{exp}\frac{n_0-n_H(\bf{r})}{n_d}\Big) \, ,
\end{equation}
where $\sigma_0$, $n_0$ and $n_d$ are the free parameters of the model.
In this work we used a 2D Gaussian distribution of hydrogen and fitted the parameters in such a way that at the point with maximum concentration of dopants, the conductivity is about $6\times10^5$ times smaller than in the pristine nickelate film.

Eq.~\ref{eq3} for potential $\varphi(\bf{r})$ was solved iteratively on a $751\times751$ 2D grid with boundary conditions $\varphi=0$ at the edges of the computational grid and at the Au contact, while a voltage $V=1$ arb. units was applied to the Pd/Pt contact.
The current $I$ was evaluated as a total current flowing in the Au contact.
In the geometry of a device with two contacts shown in Fig.~\ref{Device1}, the resistance in Fig.~\ref{Device1}a is approximately 10 times higher than in Fig.~\ref{Device1}b.
This difference is explained by redistribution of the current flow, as shown in Fig.~\ref{Device1}a-b, and the fact that even the pristine \ce{RNiO3} has low conductivity, so there is no short between the contacts.
In Fig.~\ref{Device1}(c) the gradual change of the resistance is plotted as a function of voltage pulses applied to the electrodes.
In this simulation we estimated the magnitude of each pulse to be 1\,mV with a duration of 1\,ns to match the typical values for a YBCO-based neuron described above.

\begin{figure*}
	\includegraphics[width = \linewidth]{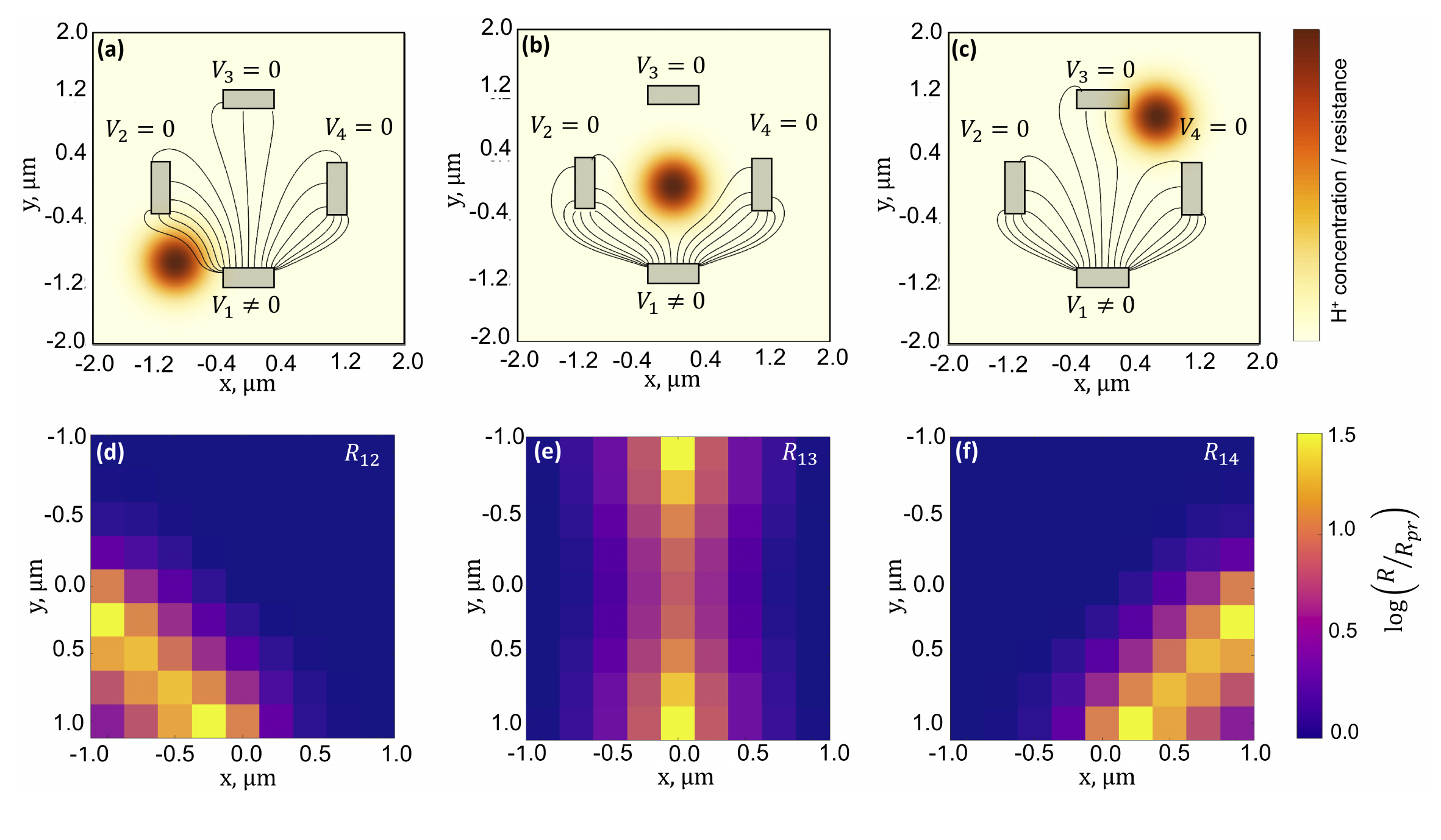}
	\caption{\label{Device2}
		(a-c) Distribution of \ce{H+} in a scheme with two overlaying pair of contacts. The current distributions are shown with black lines for three different positions of the \ce{H+} cloud between the contacts. In all cases one electrode has a positive potential $V_1\ne0$, and all other electrodes are grounded, $V_2=V_3=V_4=0$.
		(d-f) Resistance $R_{12}$ (d), $R_{13}$ (e), and $R_{14}$ (f) measured between the corresponding pairs of electrodes shown panels (a-c). Common logarithm of the ratio $R/R_{pr}$ is shown, where $R_{pr}$ is the resistance between the corresponding pair of contacts in the absence of \ce{H+} doping.
		}
\end{figure*}

At this point one can see an analogy between a magnetic vortex entering the superconducting loop of YBCO and changing its total magnetic flux, and a cloud of highly mobile \ce{H+} ions drifting between two contacts and changing the total resistance.
Another step in the creation of a disordered array of such \ce{RNiO3}-based synapses would be overlaying of contacts pairs in such a way that redistribution of \ce{H+} ions will influence the total resistance between all pairs of contacts.
An example of such a situation with two pairs of contacts is shown in Fig.~\ref{Device2}a-c.
One can place the center of the \ce{H+} distribution at any point between the four contacts.
Our computations, conducted under the same assumptions described previously, directly show the current flow between the contacts for various distributions of \ce{H+}.
The largest changes in the current flow between two contacts occur when the high resistive \ce{H+}-rich area is located directly between the contacts.
In Fig.~\ref{Device2}a-c the examples of current flow are shown for three different distributions of \ce{H+}, indicating how significantly one can change the flow by introducing the \ce{H+} ions between the pair of contacts or directly under the electrode.
In these simulations, we assumed that the voltage is applied to only one (lower) electrode, while the three other electrodes are grounded.

The resistances $R_{12}$, $R_{13}$, and $R_{14}$ between the corresponding pairs of electrodes are shown in Fig.~\ref{Device2}d-f as a function of the position of the \ce{H+} center of mass.
Each point on these plots corresponds to a certain position of the \ce{H+} cloud in the scheme with two overlaying pairs of electrodes shown in Fig.~\ref{Device2}a-c.
One can see that the resistance increases significantly when the \ce{H+} ions are concentrated not just between two electrodes, but especially close to one of them.
We should note that due to a high fourfold symmetry of the contacts, the maps of the resistance are also symmetric ($R_{13}$ has a horizontal axis of symmetry $y=0$, while $R_{12}$ and $R_{14}$ can be obtained from each other by flipping around a vertical axis $x=0$).
Similar maps can be obtained for all other pair of contacts; in the case for symmetric contacts they will look exactly like in Fig.~\ref{Device2}(d-f), but rotated over $90^{\circ}$ or $180^{\circ}$.

Overlaying of several pairs of contacts allows changing the resistance of all of them simultaneously by shifting the center of the \ce{H+} distribution.
For example, for the configuration of electrodes shown in Fig. \ref{Device2}(a-c), decreasing the synaptic weight of the vertical pair of contacts by moving \ce{H+} from point $(x,y)=(-0.5, -0.5)$ $\mu$m to $(x,y)=(0, 0)$ $\mu$m  will also cause a similar decrease of the the synaptic weight for the horizontal pair.
At the same time, the corresponding synaptic weight will be increased since the resistance between the contacts 1 and 2 will be decreased.
Obviously, the coupling between pairs of contacts can be designed to be significantly more complex by creating non-symmetric (disordered) arrangements of contacts.

This coupling between different devices on a hardware level can improve the performance of a neuromorphic circuit in two ways.
First, it creates additional connections between neurons, which can be seen as lateral connections within a single layer of an artificial network or additional feedback connections between neurons in different layers \cite{Hiratani2018, watts1998collective}.
This situation of several neurons being connected to each other via a single nanodevice with several coupled pair of contacts is similar to the disordered array of superconducting loops described earlier.
Second, allowing \ce{H+} ions to move not only between a pair of contacts, but also in orthogonal directions (so that the center of \ce{H+} distribution can be placed anywhere within a 2D device), one can create new memory states.
Each memory state corresponds to a certain distribution of the \ce{H+} and leads to a certain set of synaptic weights.
Some of the configurations may be degenerate due to any symmetry in electrode positions (as in Fig. \ref{Device2}). However this degeneracy can be removed by creating a disordered arrangement of contacts, similar to the contacts created for the superconducting loops discussed in the previous section.

In general, there are many other strongly correlated materials that display resistive properties highly sensitive to the valency of the \textit{d}-metal ion~\cite{Rosillo2020}. For instance, vanadium oxides have recently emerged as key materials for designing neuron-like resistors~\cite{delValle2019_Nature}. Manganites~\cite{Coey_manganites} and cobaltates~\cite{Galli2020} also display properties that could potentially be tuned via light-ion doping, which could then be turned into random networks. Furthermore, recent work on 5d-iridates with strong spin-orbit coupling have displayed resistive switching~\cite{Fuentes2018}, which could be used in spin-torque oscillators that also provide neuromorphic functionalities~\cite{Grollier2020}. 

\section*{Outlook and Conclusions}

A key aspect of any hardware neural network design is the connectivity between elements. The the cuprate-based loop-junction devices we describe are submicron scale and can easily be fabricated on thin films allowing for a vast array of them. Importantly, the unique aspect of utilizing similar fluxon dynamics with simple geometrical variations of superconducting loop structures to achieve both neuron and synapse operation offers a new kind of flexibility. For instance, the storage of flux quanta in superconducting loops with Josephson junctions brings about aspects such as plasticity that is required for synapses, while the oscillatory properties of these loops in response to spiking inputs resembles properties that are needed in neurons.

Furthermore, the two material platforms we present could be connected by growing an island of thin film of nickelate over the cuprate film with all the loops already patterned. Since the hydrogen is not known to affect the properties of the cuprate films, the whole device can then be implanted with hydrogen, which only affects the nickelate overlayer. Then, the two films can be wire-bonded through many possible entry paths across the networks, allowing for highly flexible and complex network design patterns. Combining the two systems opens up a particularly interesting possibility to realize a key aspect of biological brains: the ability to operate memories that span different time scales. While cuprate synapses exhibit dynamic and volatile memory with ease in “learning” and “forgetting”, the nickelate synapses exhibit long-term non-volatile memory. Therefore, the voltage spikes produced in the cuprate system can interact with the nickelate synapse arrays to update their long-term memory configurations. In summary, our proposed arrays of devices can be implemented into networks with flexible connectivity platforms.

In conclusion, for the first time, we present simulations of artificial neural network components combining superconductivity and metal-insulator transitions in complex oxides, two of the most spectacular examples of emergent physics in condensed matter.  
By creating electronic and structural disorder induced by light ions, one can design individual devices that mimic neurons and synapses in the brain. These devices can easily be combined into coupled networks using well-established lithography methods, and then the number of response states increases making them potential candidates for neuromorphic cryoelectronics. Moreover, our key finding demonstrates that a randomly spaced network of devices shows an exponential number of collective states that begins to mimic the emergent behavior of features found in living organisms. We emphasize that the use of light ions to modify the electronic response properties of oxides can  be applied to several families of strongly correlated materials, thus paving the way for a multitude of studies that can be performed in the search for a new computational paradigm. 

\acknow{We thank Ivan Schuller, Nirjhar Sarkar, Shane Cybart, Marcelo Rozenberg for fruitful discussions.
This collaborative work was supported as part of the “Quantum Materials for Energy Efficient Neuromorphic Computing” (Q-MEEN-C), an Energy Frontier Research Center funded by the U.S. Department of Energy, Office of Science, Basic Energy Sciences under Award $\#$ DE-SC0019273}

\showacknow{} 

\subsection*{Competing interests}
The authors declare no competing interests.

\end{document}